\documentclass[epj]{svjour}

\usepackage{amsmath,amssymb,graphicx}

\usepackage[applemac]{inputenc}

\newcommand{\be}{\begin{equation}}
\newcommand{\ee}{\end{equation}}
\newcommand{\bea}{\begin{eqnarray}}
\newcommand{\eea}{\end{eqnarray}}
\newcommand{\non}{\nonumber}

\begin{document}

\title{The dark aftermath of Higgs inflation}

\author{Massimiliano Rinaldi\inst{1},\inst{2} %
\thanks{\emph{E-mail:} massimiliano.rinaldi@unitn.it}
}                     % Do not remove
\institute{Namur Center for Complex Systems (naXys), University of Namur, 
Rue de Bruxelles 61, B-5000 Namur, Belgium. \and Dipartimento di Fisica \& TIFPA, Università di Trento, via Sommarive 14, 38123 Povo (TN), Italy.}
\date{Received: date / Revised version: date}

\abstract{
\noindent In this letter we study the dynamics of the late Universe when a nonminimally coupled Higgs field is present. In general, the nonminimal coupling leads to a nontrivial mixing between the gravitational degrees of freedom and the Goldstone massless bosons. We know that this is irrelevant during the inflationary phase. In contrast, in the late Universe the nonminimal coupling affects the classical equations of motion, leading to an  acceleration of the expansion rate or to a collapse that forms Q-balls. 
}

\maketitle

\section{Introduction}

\noindent The inflationary model that identifies the inflaton with the real part of the nonminimally coupled Higgs boson has gained a lot of momentum in the last years. There are many aspects that render Higgs inflation appealing. Just to name  a few,  it does not need exotic extensions of the standard model, extra dimensions or supersymmetry. In addition, there is only one parameter, namely the strength of the nonminimal coupling to gravity, which is much larger than the conformal value and it is fixed by observations through the inflationary slow-roll parameters  \cite{shapo}. Last but not least, Planck data has showed that Higgs inflation is quite a safe bet \cite{planck,chris}.

Since its formulation in Ref.\ \cite{shapo} (see also \cite{unruhetal1,unruhetal2,unruhetal3,unruhetal4}), Higgs inflation triggered many discussions especially on quantum corrections \cite{corrections1,corrections2,postma,casadio,burgessetal1,burgessetal2,barv} and on possible connections to dark energy \cite{bellido}. On a more fundamental aspect, the nonminimal coupling between Higgs field and gravity shed some light on the interplay between the mass defined in general relativity and the one coming from spontaneous symmetry breaking. In particular, the formation and the properties of ``Higgs monopoles'' were recently discussed in Ref.\ \cite{namur}.

In the original setup, Higgs inflation is based on a number of simplifying hypothesis. In particular, the unitary gauge is imposed  so that the field is a real scalar singlet \cite{shapo}. However, since the Higgs field is a complex doublet with all its component  nonminimally coupled to gravity, there is no gauge transformation able to eliminate at once all the Goldstone bosons. In fact,  these become essential to correctly compute both the classical evolution of the background \cite{kaiser1,kaiser2,kaiser3,Hertzberg} and the quantum effects  \cite{postma,casadio,burgessetal1,burgessetal2,barv} during inflation. In addition, at the energy density typical of inflation, an equivalence theorem applies such that massless Goldstone bosons are present in the spectrum \cite{pesk}. At the classical level and with nonminimal coupling, all the Goldstone bosons interact non-trivially with the gravitational field, modifying the equations of motion of the background. During inflation, the effects on the background are known to be negligible \cite{kaiser1,kaiser2,kaiser3}. However, in the late Universe this is not the case, as we are going to show in this letter.

In the next section, we review the Higgs inflationary model and we stress the role of the Goldstone bosons in the background dynamics. In section \ref{late}, we consider a simple Abelian example and show how the phase of the Higgs field leads to either the cosmic accelerated expansion or to the formation of Q-balls that might contribute to the dark matter content. We conclude in section \ref{conclusion} with few remarks on our results.

\section{Background dynamics}

\noindent  Before considering the dynamics of the  late Universe, it is instructive to review some essential features of Higgs inflation. The Lagrangian that is usually considered reads, in Jordan frame \cite{shapo} 
\bea\label{lagra}
{{\cal L}_{J}\over  \sqrt{g}}={1\over 2}(m^{2}+2\xi {\cal H}^{\dagger}{\cal H})R-(D_{\mu}{\cal H})^{\dagger}(D^{\mu}{\cal H})-V+\ldots,
\eea
where the dots stand for gauge and other standard model fields kinetic terms.
Here, $m$ is a mass parameter while $m_{p}^{2}=m^{2}+2\xi {\cal H}^{\dagger}{\cal H}$ is identified with the reduced Planck mass squared \footnote{The mass scale $m$ differs very little from the Planck mass, even during inflation \cite{shapo}.}, $R$ is the Ricci scalar, $D_{\mu}$ is the gauge covariant derivative, and ${\cal H}$ is the Higgs doublet. The dimensionless parameter $\xi$ determines the strength of the nonminimal coupling to gravity. Finally, the potential is given by 
\bea\label{pot}
V=\lambda\left({\cal H}^{\dagger}{\cal H}-{v^{2}\over 2}\right)^{2},
\eea
where $v\simeq 246$ GeV is the vacuum expectation value (vev) of the Higgs field (when there is no coupling to gravity, see below) and $\lambda\simeq 1/10$ is the self-coupling parameter. Note that the term ${\cal H}^{\dagger}{\cal H}R$ is generically necessary to eliminate the divergences of the stress tensor in the semi-classical approximation \cite{callan}.

The inflationary behaviour is manifest in the Einstein frame, which is related to the Jordan frame by the conformal transformation $\bar g_{\mu\nu}=\Omega^{2}g_{\mu\nu}$, where $\Omega^{2}=1+2\xi {\cal H}^{\dagger}{\cal H}/m^{2}$. In this frame one has 
\bea\label{eframe}
{{\cal L}_{E}\over \sqrt{\bar g}}={m^{2}\over 2}\bar R-{1\over \Omega^{2}}(\bar D_{\mu}{\cal H})^{\dagger}(\bar D^{\mu}{\cal H})-{V\over \Omega^{4}}-{3\xi^{2}\over m^{2}\Omega^{4}}\bar g^{\mu\nu}\partial_{\mu}({\cal H}^{\dagger}{\cal H})\partial_{\nu}({\cal H}^{\dagger}{\cal H})+\ldots
\eea
One way  to write this Lagrangian in canonical form is to adopt the unitary gauge, which assumes that ${\cal H}^{T}(x)=(0,h(x))/\sqrt{2}$, and to define the new field $\chi$ by the relation
\bea\label{hchi}
{d\chi\over dh}=\Omega^{-2}\sqrt{\Omega^{2}+6\xi(\Omega^{2}-1)},
\eea
so that
\bea
{{\cal L}_{E}\over  \sqrt{\bar g}}={m^{2}\over 2}\bar R-{1\over 2}\bar g^{\mu\nu}\partial_{\mu}\chi\partial_{\nu}\chi-U(\chi)+\ldots.
\eea
At high energy, $h\gg m/\sqrt{\xi}$,  the potential $U(\chi)$ is sufficiently flat to allow for a slow-rolling phase of the field $\chi$.  The requirement that inflation lasts long enough to satisfy the current observational data fixes the coupling constant to $\xi\simeq 5\times 10^{4}\sqrt{\lambda}$, although some running occurs because of loop corrections \cite{barv}. At the end of inflation the energy decreases until $h\simeq \chi$, the two frames become indistinguishable, and the potential takes the usual form $(\lambda/4)(h^{2}-v^{2})^{2}$. 

The choice of the unitary gauge usually goes along with the possibility of expanding the Lagrangian around a classical, time-independent vacuum state. However, when the Higgs field components are nonminimally coupled to gravity, the classical vacuum state is not constant in time. This can be easily seen by writing down the Klein-Gordon equation from the Lagrangian \eqref{lagra}, which has a term proportional to $\xi R {\cal H}$. Since $\xi\neq 0$, $|{\cal H}|=v$ is not a solution \footnote{This fact is essential for the existence of the Higgs monopole, see \cite{namur}.}, see also Eq.\ \eqref{kg} below. At the best, the vev of the theory can be expressed as a series of nonlocal operators acting on powers of the Ricci scalar \cite{shapiro}, explicitly calculable only on Einstein spaces \cite{madsen}. As a consequence, we cannot expand the Lagrangian at $|{\cal H}|=v$, but only around some time-dependent classical background solution ${\cal H}_{\rm cl}(t)$ (which might asymptotically coincide with $v$, like in the cosmological case as we will show in the next section). Such expansion involves also metric fluctuations, precisely because of the nonminimal coupling with all the components of the Higgs doublet in Jordan frame, see e.g. \cite{casadio}. In Einstein frame, the coupling is minimal but the kinetic terms of the Higgs component are not canonical and their expansion involves again gravitational and bosonic degrees of freedom \cite{kaiser1,kaiser2,kaiser3}. In fact, the right canonical fields to be used for quantization should be the Mukhanov variables \cite{mukhanov}, which mix scalar and tensor field fluctuations and yield unambiguously the correct particle content. The unitary gauge, which acts only scalar and gauge fields, cannot cancel the mixing between gravitational field fluctuations and Goldstone bosons, which acquire a dynamical role also at the classical level. For this reason, Higgs inflation should be considered as a multifield model of inflation, as pointed out in \cite{kaiser1,kaiser2,kaiser3,Hertzberg}. Although it was shown that the influence of the Goldstone bosons is negligible on the classical inflationary dynamics, certain observables are sensitive to the multifield nature of Higgs inflation and future observation might be able to discriminate among various configurations in the parameter space \cite{kaiser1,kaiser2,kaiser3}. We now show that, during the late-time evolution of the Universe, the influence of the Goldstone bosons can be important also at the classical level.

To illustrate this effect, let us assume, for simplicity, a $U(1)$ gauge symmetry and  that, on a cosmological background, the Higgs field has the form ${\cal H}(t)=h(t)/\sqrt{2}\,\exp(i\theta(t))$. After the same conformal transformation as above, the Lagrangian in Einstein frame reads
\bea
{{\cal L}_{E}\over  \sqrt{\bar g}}={m^{2}\over 2}\bar R-{1\over 2\Omega^{4}}\left[\Omega^{2}+6\xi (\Omega^{2}-1)\right](\bar\partial_{\mu}h)^{2} -{h^{2}\over 2\Omega^{2}} (\bar\partial_{\mu} \theta)^{2}-{V(h)\over \Omega^{4}},
\eea
where we omitted the gauge fields. We assume that, being conformally invariant, gauge fields are washed out by the inflationary expansion, even in the case when they become massive \cite{mukha}.

During inflation, we have $h/\Omega\simeq 1$  and the dynamics is fully driven by the field $\chi$, defined in terms of $h$ by Eq.\ \eqref{hchi}, which  slowly rolls down along the potential \cite{shapo}
\bea
U(\chi)\simeq {\lambda m^{2}\over 4\xi^{2}}\left[1+\exp\left(-{2\chi\over \sqrt{6}m}\right)\right]^{-2}.
\eea
The phase term is decoupled from both $h$ and the gravity sector so it does not play any role in the inflationary dynamics. In fact, its equation of motion is simply $\bar \square \theta=0$, which gives $\theta \sim \exp{(-3Ht)}$, where $H$ is the Hubble parameter (that is almost constant during inflation). 

The situation changes  at the end of inflation, when the energy density drops, $\Omega\rightarrow 1$, and $h\rightarrow \chi$. The low-energy Lagrangian reduces to
\bea\label{lowlagra}
{{\cal L}_{E}\over  \sqrt{\bar g}}={m^{2}\over 2}\bar R-{1\over 2}(\bar\partial_{\mu}\chi)^{2} -{\chi^{2}\over 2} (\bar\partial_{\mu} \theta)^{2}-{\lambda\over 4}(\chi^{2}-v^{2})^{2},
\eea
and the field $\theta$ clearly has a non-trivial dynamics. In fact, this model closely reminds the ``spintessence'' presented in Ref.\ \cite{spintessence}, and  we now show how it can lead to either dark energy or dark matter.

\section{Late-time evolution}\label{late}

\noindent  As a first step, we investigate whether a time-dependent $\theta$  can accelerate the current Universe.  The equations of motion derived  from the Lagrangian  \eqref{lowlagra} read
\bea\label{hubble}
H^{2}={\kappa^{2}\over 3}\left({\dot \chi^{2}\over 2}+{{\cal Q}^{2}\over 2\chi^{2}a^{6}}+V+\rho_{m}\right),\\\dot H=-{\kappa^{2}\over 2}\left[\dot \chi^{2}+{{\cal Q}^{2}\over \chi^{2}a^{6}}+\rho_{m}(1+\omega_{m})\right],\\
\ddot\chi+3H\dot\chi-{{\cal Q}^{2}\over \chi^{3}a^{6}}+{dV\over d\chi}=0\ .\label{kg}
\eea
Here, $H=\dot a/a$ is the Hubble parameter, $V=(\lambda/ 4)(\chi^{2}-v^{2})^{2}$, and $\rho_{m}$ is the energy density of a barotropic fluid with $\omega_{m}=0$ for dust and $\omega_{m}=1/3$ for radiation, and whose evolution is determined by the usual equation of state $\dot\rho_{m}+3H\rho_{m}(\omega_{m}+1)=0$. ${\cal Q}$ is the conserved charge associated to $\theta$ through the equation $\dot \theta={\cal Q}/(\chi^{2}a^{3})$. As mentioned above,  $\chi(t)=v$ is not a  solution of the Klein-Gordon equation when ${\cal Q}\neq 0$ (and $\chi$ is finite).  To see why this leads to acceleration, we write the above equations in terms of the first derivative of $N=\ln a$, through the change of variables  $x=\kappa\dot \chi/(\sqrt{6}H)$, $y=\kappa\sqrt{V}/(\sqrt{3}H)$, $z=\kappa {\cal Q}/(\sqrt{6}\chi a^{3}H)$, $w=\kappa\chi/\sqrt{6}$, and $L=-(1/\kappa)(d\ln V/d\chi)$. The equations of motion are then written as a dynamical system that generalizes the standard one \cite{DEbook} by adding the new variable $z$, that is
\bea\label{sys}
{dx\over dN}&=&-3x+{z^{2}\over w}+\sqrt{3\over 2}Ly^{2}+{3\over 2}xP,\\
{dy\over dN}&=&-\sqrt{3\over 2}Lxy+{3\over 2}yP,\\
 {dz\over dN}&=&-3z-{zx\over w}+{3\over 2}zP,\\
{dw\over dN}&=&x,
\eea
 where $P=(1-\omega_{m})(x^{2}+z^{2})+(1+\omega_{m})(1-y^{2})$. The Hamiltonian constraint following from Eq.\ \eqref{hubble} implies that
\bea
{\kappa^{2}\rho_{m}\over 3H^{2}}=1-x^{2}-y^{2}-z^{2},
\eea
while the parameter of the equation of state for the scalar field reads
\bea\label{omegachi}
\omega_{\chi}={x^{2}+z^{2}-y^{2}\over x^{2}+z^{2}+y^{2}}.
\eea
The Hamiltonian constraint can be written as $\Omega_{m}+\Omega_{\chi}=1$, where $\Omega_{m}=\kappa^{2}\rho_{m}/(3H^{2})$ is the energy density of matter or radiation and $\Omega_{\chi}=x^{2}+y^{2}+z^{2}$ is the one associated to the scalar field. The system has only one fixed point at  $(x,y,z,w)=(0,0,0,\kappa v/\sqrt{6})$ that turns out to be a \emph{saddle point} for all $\omega_{m}$ \footnote{Actually, the fixed point is $(x,y,z,w)=(0,0,0,{\rm any})$ but $w$ and $y$ are not independent and $y=0$ implies $\chi=v$ and $w=\kappa v/\sqrt{6}$.}.  This means that a Universe dominated by  matter or radiation only (i.e. such that $\Omega_{\chi}=0$ and $V=0$) is an unstable equilibrium point in the phase space. 
The deceleration parameters $q=-1-\dot H/H^{2}$ in the new variables reads
\bea\label{q}
q=-1+{3\over 2}(x^{2}+y^{2})(1-\omega_{m})-{3\over 2}(1+\omega_{m})y^{2}.
\eea
In the vicinity of the fixed point we find that $x,y\sim e^{-3(1-\omega_{m})N/2}$ while $z\sim e^{3(1+\omega_{m})N/2}$. Therefore, after a sufficient number of e-folds, and for any $0<\omega_{m}<1$, the last term in Eq.\ \eqref{q} will always take over on the second, leading to a negative deceleration parameter and  an accelerated expansion.

The fact that the charge ${\cal Q}$ does not vanish, no matter how slowly the phase $\theta$ changes, is crucial. When ${\cal Q}=0$, the $z$-equation of the system \eqref{sys} is trivial and  two more fixed points appear at $(x,y,w)=(0,\pm 1,0)$. These correspond to a Universe dominated by the scalar field and to an equation of state parameter $\omega_{\chi}=-1$. However, these are saddle points, thus the Universe evolves away from a dark energy dominated phase. 

\begin{figure*}
\begin{center}
	\includegraphics[scale=0.45]{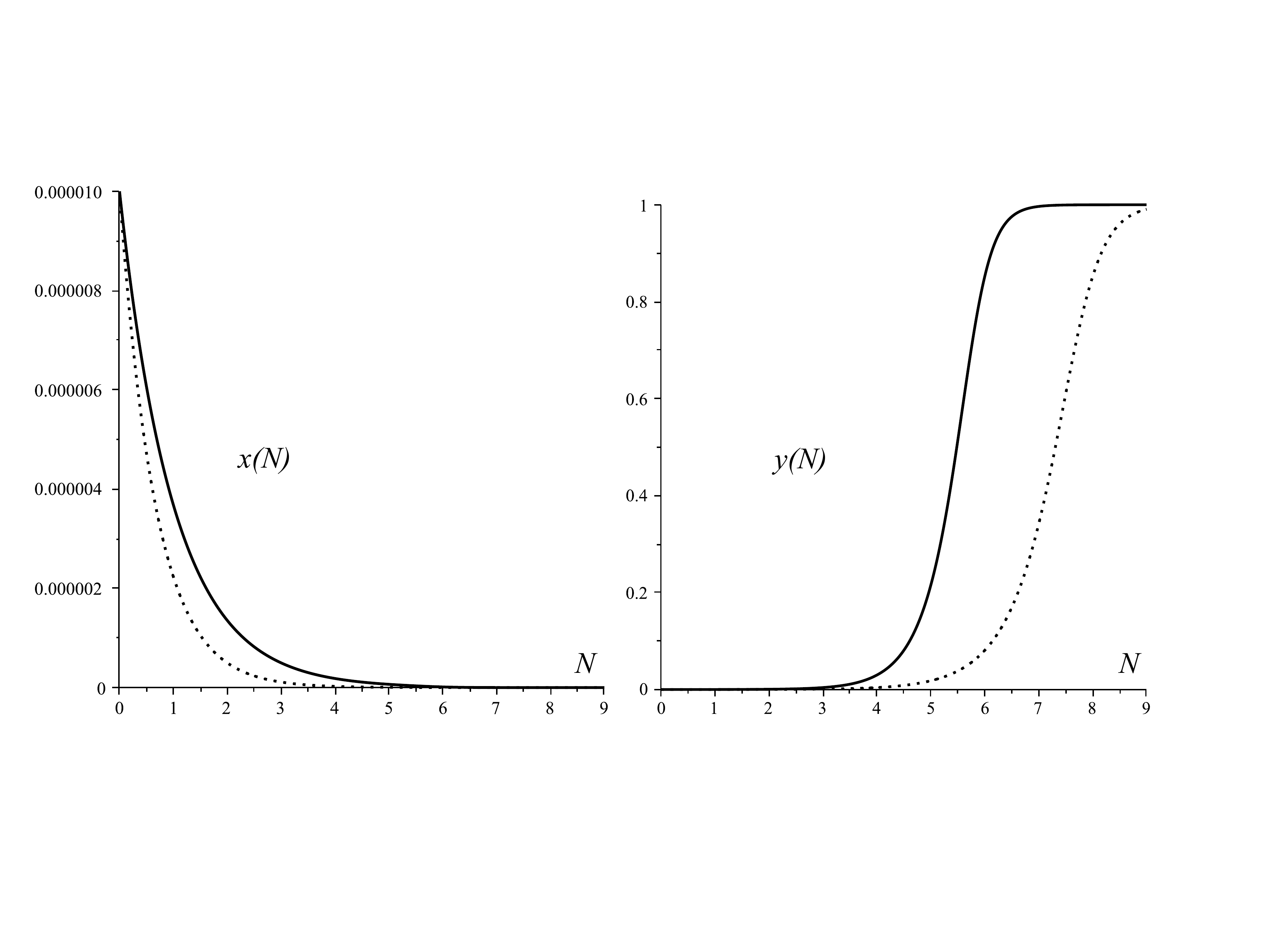}
\end{center}
\caption{Plot of $x(N)$ (on the left) and of $y(N)$ (on the right) for $\omega=0$ (dotted) and $\omega=1/3$ (solid). The initial conditions are $x(0)=y(0)=z(0)=10^{-3}$ and $w(0)=\kappa v(1+10^{-5})/\sqrt{6}$. The plot of $z(N)$ is almost identical to the one of $x(N)$.}\label{fig1} 
\end{figure*}

\begin{figure*}
\begin{center}
	\includegraphics[scale=0.62]{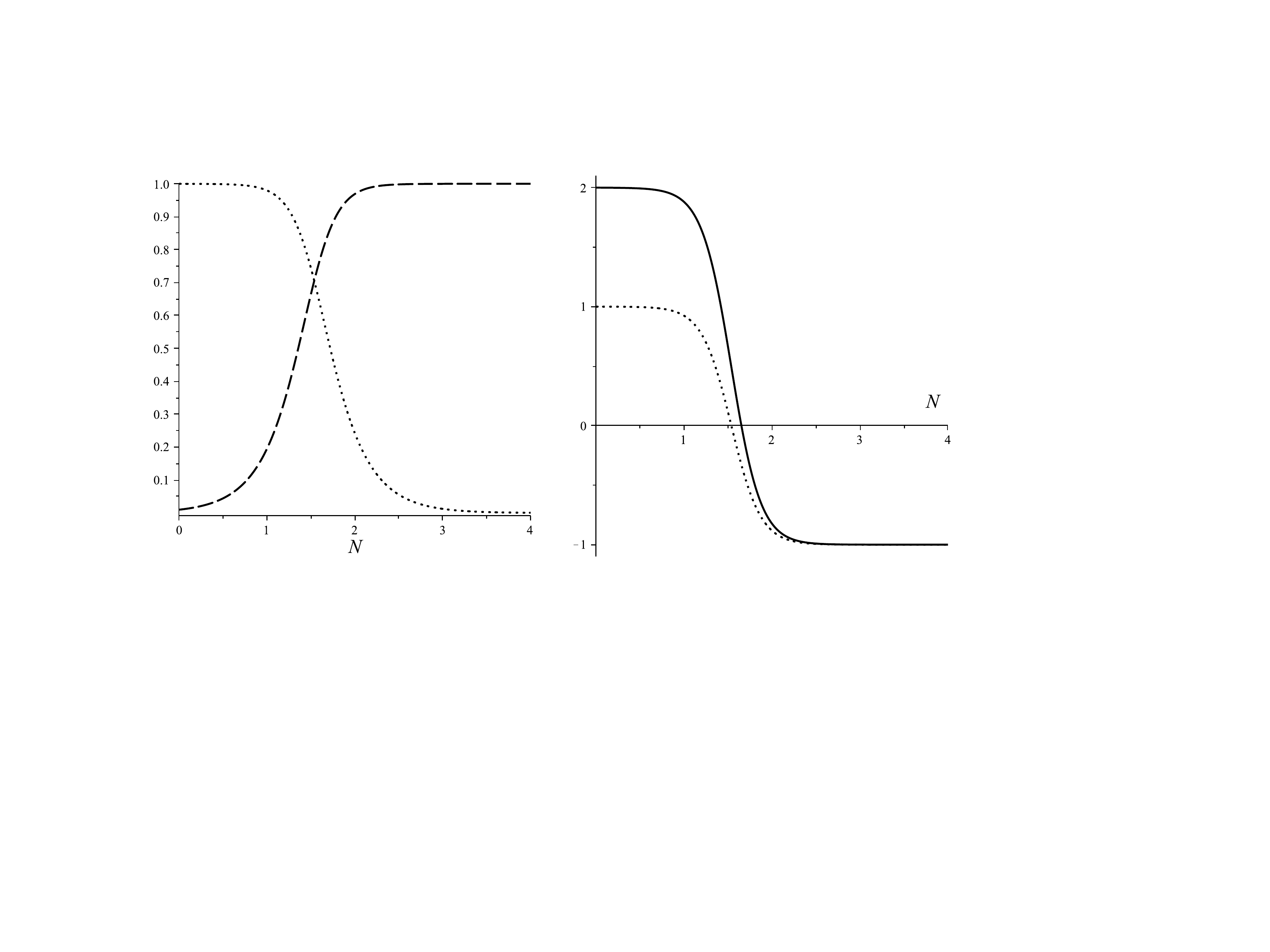}
\end{center}
\caption{On the left: plot of $x(N)$ (dotted line) and of $y(N)$ (dashed line) for $\omega_{m}=0$. On the right: plot of $\omega_{\chi}(N)$ (dotted line) and $q(N)$ (solid line). The initial conditions are: $x(0)=0.9999$, $y(0)=0.0099$, $z(0)=0.0001$, and $w(0)=11\kappa v/\sqrt{6}$. }\label{fig2} 
\end{figure*}

To clarify further our findings, we plotted the solutions to the system \eqref{sys} in the case of $\omega=0$ and $\omega=1/3$ when the initial conditions are close to the fixed point, see Fig. \ref{fig1}. In both cases, we see that $x(N)$ (together with $z(N)$) tends to zero while $y(N)$ tends to one and, from Eq.\ \eqref{omegachi}, that $\omega_{\chi}$ becomes closer and closer to $-1$. We also plotted the functions $x(N)$ $y(N)$ ($z(N)$ has much smaller values than the other two but it is never vanishing, as it should be) with initial conditions typical of the matter-radiation equality era, see the left plot of Fig.\ \ref{fig2}. Even in this case, the system evolves towards a dark energy dominated phase. We tried several different initial conditions and we found that the results do not change qualitatively. On the right of  Fig.\ \ref{fig2} we plotted the deceleration parameter and $\omega_{\chi}$ and we see that they both converge towards $-1$, indicating late time acceleration.

From these plots it is also apparent that the system seems to evolve towards the points $(x,y,z,w)=(0,\pm 1,0,{\rm any})$ although these are not fixed points. The reason is that in the regime where $\chi $ is close to $v$, $L$ is very small. If we set $L=0$ in the system \eqref{sys} we find that $(x,y,z,w)=(0,\pm 1,0,{\rm any})$ are two stable fixed points. This further support the result that the late Universe is gradually dominated by the scalar field and accelerated, no matter the initial conditions. In few words, by chasing the vacuum state $\langle \chi \rangle=v$, the Higgs field ends up by dominating the dynamics of the late Universe.

%%%%% FIG 1 %%%%%%
%%%%% %%%%%%%%%%

%%%%% FIG 2 %%%%%%
%%%%% %%%%%%%%%%

The simple considerations made so far hide a caveat that might change completely the fate of the Higgs field at late time. In fact, as noted in Ref.\ \cite{spintessence} (see also Ref.\ \cite{kasuya1,kasuya2}), the cosmological equations of motion of spintessence are generically unstable and the fluctuations of the field $\theta$ can grow exponentially and collapse into ``Q-balls'', namely solitonic extended objects described many years ago by Coleman in Ref.\ \cite{coleman}. If this is the case, Q-balls should be considered as dark matter rather than dark energy because the equation of state parameter vanishes. Whether or not this happens depends upon the magnitude of the angular velocity $\dot\theta$ and on the shape of the potential. Let us consider the perturbed cosmological metric $ds^{2}=-(1+2\Phi)dt^{2}+(1-2\Phi)a^{2}\delta_{ij}dx^{i}dx^{j}$, where $\Phi=\Phi(t,\vec x)$, and replace the field $\chi$ and $\theta$ by respectively $\chi(t)+\delta\chi(t,\vec x)$ and $\theta(t)+\delta\theta(t,\vec x)$ in the Lagrangian \eqref{lowlagra}, which takes the form ${\cal L}_{E}^{0}+{\cal L}_{E}^{2}$. The second term is quadratic in the fields $\delta\chi$, $\delta\theta$ and $\Phi$ and its variation with respect to these yields, respectively, the equations of motion
\bea
&&\ddot{\delta\chi}+3H\dot{\delta\chi}+\left(V''-\dot\theta^{2}-a^{-2}\nabla^{2}\right){\delta\chi}=\\\non
&&4\dot\chi\dot\Phi-2\Phi V'+2\chi\dot\theta\dot{\delta\theta},\\
&&\ddot{\delta\theta}+3H\dot{\delta\theta}- a^{-2}\nabla^{2}{\delta\theta}=\\\non
&&4\dot\theta\dot\Phi-2{\dot\delta\chi\over \chi}\dot\theta+2{\dot\chi\over\chi}\left({\delta\chi\over\chi}\dot\theta-\dot{\delta\theta}\right),\\
&&a^{-2}\nabla^{2}\Phi-3H\dot\Phi-3H^{2}\Phi=\\\non
&&{\kappa^{2}\over 2}\left[ \dot\chi\dot{\delta\chi}+V'\delta\chi+\chi^{2}\dot\theta\dot{\delta\theta}+\chi\dot\theta^{2}\delta\chi-\Phi(\dot\chi^{2}+\chi^{2}\dot\theta^{2}) \right].
\eea
To study the stability we set $\delta\chi=\delta\chi_{0}\exp (\omega t+i\vec k\cdot\vec x)$ $\delta\theta=\delta\theta_{0}\exp (\omega t+i\vec k\cdot\vec x)$ and $\Phi=\Phi_{0}\exp (\omega t+i\vec k\cdot\vec x)$ and find the critical Jeans value $k_{J}$ such that $\omega=0$. We find that, for $\chi\simeq v$,
\bea
k_{J}^{2}\simeq 8\lambda v^2 \kappa^2(\chi^{2}-v^{2})+{\cal O}((\chi^{2}-v^{2})^{2}),
\eea
and that for $k<k_{J}$, $\omega^{2}>0$, i.e. the fluctuations grow exponentially. From the discussion above we know that $\chi$ cannot coincide with $v$ even in the present Universe, therefore there always exist an instability band in the fluctuation spectrum. It is interesting to see how the critical wavelength associated to $k_{J}$ is deeply related to the parameters of the standard model $v$ and $\lambda$. In conclusion, the formation of Q-balls is possible for sufficiently large wavelength and this leaves also the possibility that the same field is responsible for both dark matter and dark energy, according to the wave number of its fluctuations \footnote{In principle, Q-balls with a local gauge symmetry are unstable \cite{coleman}. However, if the gauge coupling and the $U(1)$ charge are not too large, stability can be achieved, see \cite{lee}.}. In the case of larger symmetry groups it may happen that some Goldstone bosons are responsible for dark matter and the remaining ones for dark energy.

\section{Conclusions}\label{conclusion}

\noindent  A detailed analysis of the cosmological evolution requires a numerical study of the equations that goes beyond the scope of this letter and will be presented elsewhere. The main point is however clear: a nonminimally coupled Higgs field  can generate dark energy and/or dark matter, provided one keeps track of the dynamical evolution of the associated Goldstone bosons. In this letter we considered the simplest symmetry $U(1)$ and only one nonvanishig Goldstone boson but the generalization to larger symmetry groups and to multiple bosons should be straightforward \cite{ONsym}.  

We finally remark that our findings are independent of the value of the nonminimal coupling parameter $\xi$. In fact, what we found is a low-energy effect that takes place whenever we nonminimally couple gravity to a complex scalar field governed by the potential \eqref{pot}. In other words, Higgs inflation is not strictly required as well as the Higgs field used here does not necessarily coincide with the standard model one: the quartic potential yields the acceleration but the value of $v$ is in fact irrelevant. What really matters is the displacement of the value of the Higgs field in the classical vacuum state from $v$. However, as parsimony and simplicity are guiding principles of physical sciences, the idea that the standard model Higgs field is responsible for both inflation and dark ages is very tempting.

\section*{Acknowledgments}

\noindent This work is partially funded by the ARC convention No.11/15-040. We thank A.\ F\"uzfa, R.\ Durrer, C.\ Ringeval, R.\ Casadio for discussions, D.\ Kaiser for valuable correspondence,  and E.\ Rametta for support.

\end{document}